\documentclass[aps,prd,preprintnumbers,preprint]{revtex4-1}

\usepackage{amsmath,amssymb,bm,bbm}
\usepackage{graphicx,graphics,color}
\usepackage[colorlinks=true,urlcolor=blue,citecolor=blue,anchorcolor=blue]{hyperref}
\usepackage{dsfont}
\usepackage{setspace}
\usepackage{ulem}

\newcommand{\fref}[1]{Fig.~\ref{f.#1}}

\newcommand{\eref}[1]{Eq.~(\ref{e.#1})}

\newcommand{\tref}[1]{Table~\ref{t.#1}}

\usepackage[dvipsnames]{xcolor}

\usepackage{soul}

\begin{document}

\preprint{JLAB-THY-20-3257}

\title{Confronting lattice parton distributions with global QCD analysis}

\author{J.~Bringewatt}
\affiliation{\mbox{Department of Physics, University of Maryland, College Park, Maryland 20742, USA}}
\author{N.~Sato}
\affiliation{Jefferson Lab, Newport News, Virginia 23606, USA}
\author{W.~Melnitchouk}
\affiliation{Jefferson Lab, Newport News, Virginia 23606, USA}
\author{Jian-Wei Qiu}
\affiliation{Jefferson Lab, Newport News, Virginia 23606, USA}
\author{F.~Steffens}
\affiliation{\mbox{Instit\"ut f\"ur Strahlen- und Kernphysik, Universit\"at Bonn, 53115 Bonn, Germany}}
\author{M.~Constantinou}
\affiliation{\mbox{Department of Physics, Temple University, Philadelphia, Pennsylvania 19122, USA} \\ \\
{\bf Jefferson Lab Angular Momentum (JAM) Collaboration \\ 
}}

\begin{abstract}
We present the first Monte Carlo based global QCD analysis of spin-averaged and spin-dependent parton distribution functions (PDFs) that includes nucleon isovector matrix elements in coordinate space from lattice QCD.
We investigate the degree of universality of the extracted PDFs when the lattice and experimental data are treated under the same conditions within the Bayesian likelihood analysis.  
For the unpolarized sector, we find rather weak constraints from the current lattice data on the phenomenological PDFs, and difficulties in describing the lattice matrix elements at large spatial distances. 
In contrast, for the polarized PDFs we find good agreement between experiment and lattice data, with the latter providing significant constraints on the spin-dependent isovector quark and antiquark distributions.
\end{abstract}

\date{\today}
\maketitle

\section{Introduction}

Recent years have witnessed remarkable advances in the extraction of light-cone parton distributions from first-principles lattice QCD calculations of nucleon matrix elements of nonlocal operators.
The standard paradigm for determining parton distribution functions (PDFs) empirically has involved factorizing high-energy observables into calculable, process-dependent hard-scattering cross sections and nonperturbative functions characterizing the intrinsic properties of the nucleon.
On the other hand, lattice QCD allows the nonperturbative nucleon structure to be computed directly, offering a complementary path, along with the phenomenological efforts, to mapping out the structures in the nucleon that emerge from the quark and gluon degrees of freedom in QCD.

While PDFs cannot be calculated directly in lattice QCD, there are novel methods that give access to the $x$ dependence of PDFs (for a recent review see Ref.~\cite{Cichy:2018mum}).
One of the most notable approaches is that via parton quasi-distributions \cite{Ji:2013dva}, which allow one to access PDFs via matrix elements of momentum-boosted hadrons coupled with nonlocal spatial operators.
For finite, but large enough, momenta the Fourier transforms of these matrix elements are related to the light-cone PDFs by applying a matching procedure using Large Momentum Effective Theory (LaMET)~\cite{Ji:2014gla}. 
In the quasi-PDF approach, the connection between the lattice data and the PDFs is not direct, however.
Firstly, the relation involves a Fourier transform from lattice coordinate space into momentum space.
More significantly, it requires a nontrivial perturbative matching, up to power corrections, to relate the light-cone correlation functions encoded in the definition of the PDFs with the off-light-cone correlation functions accessible in the lattice calculations.

The traditional method for reconstructing the $x$-dependence of light-cone PDFs from lattice data has involved performing a truncated discrete Fourier transform (DFT) of the calculated lattice matrix elements.
However, the truncation of the Fourier series restricts the information that can be inferred about the PDFs, as only the region measured by the lattice data in coordinate space propagates into the extracted PDFs.
This procedure invariably limits the possibility of directly incorporating additional information from experimental data, and ultimately the ability to confront the universality of the PDFs across lattice and experimental observables. 
Currently, various advanced reconstruction methods are being investigated~\cite{Karpie:2019eiq}.

An alternative approach to this problem is to Fourier transform the matching between quasi-PDFs and PDFs into coordinate space, where the hadronic matrix elements are calculated in lattice QCD.
Such direct matching between the coordinate space matrix elements, which define the quasi-PDFs, and the light-cone PDFs was actually proved to be valid to all orders in QCD perturbation theory when the nonlocal spatial operators 
are sufficiently localized into a small distance~\cite{Ma:2014jla, Ma:2017pxb, Izubuchi:2018srq}. 
This approach allows lattice data to be treated on an equal footing with the experimental observables, and fitted simultaneously within the same analysis.

In this paper, we adopt this approach to perform the first Monte Carlo (MC) based global QCD analysis of both spin-averaged and spin-dependent PDFs from unpolarized and polarized DIS, Drell-Yan, and lattice data~\cite{Alexandrou:2018pbm}.
For the analysis we employ the theoretical framework of the Jefferson Lab Angular Momentum (JAM) Collaboration~\cite{JAM15, JAM17, JAM19}, using MC Bayesian inference sampling methodology, parametrizing the PDFs at the input scale $\mu$, and solving the evolution equations in Mellin space to relate different observables measured at different scales.
An analysis for the spin-averaged PDFs using lattice data within the NNPDF framework can be found in Ref.~\cite{Cichy:2019ebf}.

While the interpretability of the lattice data in terms of PDFs is subject to many systematic uncertainties, including discretization effects, finite volume effects, higher order perturbative corrections in the matching kernel, and power corrections, the inferred PDFs from the DFT exhibit promising general trends consistent with phenomenological distributions, but also some unexpected features in the sea quark sector.
In particular, the $\bar{d} - \bar{u}$ antiquark asymmetry at intermediate parton momentum fractions $x \sim 0.1$ is found to have the opposite sign in the lattice simulations compared to that in phenomenological analyses.
Similarly, for the spin-dependent PDFs there are indications from some global QCD studies that the $\Delta\bar{u} - \Delta\bar{d}$ asymmetry may also be opposite in sign to that from the lattice DFT results, although the uncertainties here are considerably larger.

Although the lattice systematic effects still need to be fully understood, it is nevertheless important to explore these intriguing features by confronting the lattice results with experimental data in the framework of QCD global analysis.
Such an analysis can aim to achieve three goals:
firstly, test the feasibility of including lattice data with experimental data using continuous Fourier transform from PDFs to lattice matrix elements; 
secondly, establish the degree of universality of the PDFs across the data space; 
and finally, identify the kinematic regions where the lattice data are compatible with the experimental measurements.

In proceeding with this narrative, we begin in Sec.~\ref{s.methodology} by describing the theoretical framework for the lattice matrix elements and experimental cross sections to be used in the global analysis.
In Sec.~\ref{s.bayesian} we discuss the Bayesian methodology used in previous JAM global QCD analyses to make inferences on PDFs from the data.
The numerical results from our global analysis are presented in Sec.~\ref{s.analysis}, for both the spin-averaged and spin-dependent PDFs, where we assess the impact of the lattice data on the QCD analysis of the experimental cross sections, and contrast the PDFs with those extracted using the DFT method.
We in addition compare the results of the first few moments of the unpolarized and helicity PDFs with and without the lattice data, and study the relative importance of different of coordinate space on the PDF determination.
Finally, concluding remarks and suggestions for further research are summarized in Sec.~\ref{s.conclusions}.

\section{Lattice Methodology}
\label{s.methodology}

In this analysis, we consider the nucleon matrix elements of nonlocal operators, for a given Dirac structure $\Gamma$, defined by
\begin{align}
    &{\cal M}_{[\Gamma]}^q(z,\mu) =  Z_\Gamma(z,\mu)\,
\langle N(P_3) \vert\,
    \overline\psi_q(0,z)\, \Gamma\,  W_3(z)\, \psi_q(0,0)\,
    \vert N(P_3) \rangle\,,
\label{e.ME}
\end{align}
where $\vert N(P_3) \rangle$ is the nucleon state with longitudinal momentum $P_3$, and $\psi_q$ is the quark field operator for flavor $q$.
The quark field operators, which have been shown to be multiplicatively renormalizable \cite{Ishikawa:2017faj, Ji:2017oey}, are displaced along the third spatial component in coordinate space with length $z$, and connected by a straight Wilson line $W_3(z)$.
The point-by-point multiplicative renormalization factor $Z_\Gamma$ is included to renormalize the field operators, as well as the power-law divergence associated with the Wilson line, using a prescription developed in Refs.~\cite{Alexandrou:2017huk, Constantinou:2017sej}.
The prescription is based on an RI-type scheme and is applied for each value of $z$ separately, with $Z_\Gamma(z,\mu)$ converted to the $\overline{\rm MS}$ scheme and evolved to a scale $\mu=2$~GeV.
Other prescriptions for the point-by-point multiplicative renormalization were also proposed in Refs.~\cite{Orginos:2017kos, Braun:2018brg, Chen:2017mzz, Li:2020xml}.
The Dirac structure $\Gamma$ allows us to access spin-averaged distributions using $\Gamma = 
\gamma^0
$, or spin-dependent distributions by taking $\Gamma = 
\gamma^3 \gamma_5
$.
In practice, the lattice data are available for the isovector flavor combination $q = u-d$, to which we will restrict our analysis.

The spin-averaged and spin-dependent quasi-PDFs are defined in terms of Fourier transforms of the renormalized matrix elements
    ${\cal M}_q \equiv {\cal M}_{[\gamma^0]}^q$ and
    ${\cal M}_{\Delta q} \equiv {\cal M}_{[\gamma^3 \gamma_5]}^q$,
respectively, with respect to the length of the Wilson line, $z$, by
\begin{subequations}
\begin{align}
\widetilde f_q\big(x,\mu,P_3\big)
&= 
P_3
\int_{-\infty}^{\infty}
   \frac{dz}{
2\pi}\, e^{ix P_3 z}\,
   {\cal M}_q(z,\mu),        \\
\Delta \widetilde f_q\big(x,\mu,P_3\big)
&=  
P_3
\int_{-\infty}^{\infty}
   \frac{dz}{
 2\pi}\, e^{ix P_3z}\,
   {\cal M}_{\Delta q}(z,\mu) \, .
 \end{align}
\end{subequations}
The quasi-PDFs are related to the usual light-cone PDFs via a perturbative matching procedure in LaMET~\cite{Ji:2013dva, Ji:2014gla},
\begin{subequations}
\label{eq.lc-to-qpdf}
\begin{align}
{\widetilde f}_q\big(x,\mu,P_3\big)
&= \int_{-1}^{1} \frac{d\xi}{|\xi|}\,
    C_q\Big( \frac{x}{\xi}, \frac{\mu}{\xi P_3} \Big)\,
    f_q(\xi,\mu)
 + \mathcal{O}\bigg(\frac{m^2}{P_3^2},\frac{\Lambda_{\mbox{\tiny QCD}}^2}{x^2P_3^2}\bigg),     \\
\Delta {\widetilde f}_q\big(x,\mu,P_3\big)
&= \int_{-1}^{1} \frac{d\xi}{|\xi|}\,
    C_{\Delta q}\Big( \frac{x}{\xi},\frac{\mu}{\xi P_3} \Big)\, 
    \Delta f_q(\xi,\mu)
 + \mathcal{O}\bigg(\frac{m^2}{P_3^2},\frac{\Lambda_{\mbox{\tiny QCD}}^2}{x^2P_3^2}\bigg),
\end{align}
\end{subequations}
where the light-cone PDFs $f^q$ and $\Delta f^q$ have support in the range $x \in[-1,1]$.
The matching kernels $C_q$ and $C_{\Delta q}$ are calculable in perturbative QCD, and in our analysis we use the 1-loop expression from Ref.~\cite{Alexandrou:2019lfo}, which ensures valence quark number conservation.
To this order, for the nonsinglet combination $u-d$ the spin-averaged and spin-dependent 
kernels are equivalent, $C_q = C_{\Delta q}$.
The conservation of valence quark number is achieved by a modification outside the physical region, bringing the quasi-PDFs in the modified $\overline{\rm MS}$\, (${\rm M}\overline{\rm MS}$) scheme.
This change also requires the conversion of the renormalization factor $Z_\Gamma(z,\mu)$ to the ${\rm M}\overline{\rm MS}$. 
Note that the light-cone PDFs are in the standard $\overline{\rm MS}$ scheme.
The corrections to the relations in (\ref{eq.lc-to-qpdf}) are of order ${\cal O}(m^2/P_3^2)$, where $m$ represents any hadron or quark mass scale, and ${\cal O}(\Lambda_{\mbox{\tiny QCD}}^2/x^2 P_3^2)$, where $\Lambda_{\mbox{\tiny QCD}}$ is the QCD scale parameter.

An alternative way to express the relation between the matrix elements and the light-cone PDFs is through the use of the inverse Fourier transform,
\begin{subequations}
\label{e.M}
\begin{align}
{\cal M}_q(z,\mu) 
  = &\int_{-\infty}^{\infty} dx\, e^{-i x P_3 z}
    \int_{-1}^{1} \frac{d\xi}{|\xi|}\, 
    C_q\Big(\frac{x}{\xi},\frac{\mu}{\xi P_3}\Big)\,
    f_q(\xi,\mu)\ +\ \cdots\, ,                    \\
{\cal M}_{\Delta q}(z,\mu) 
  = &\int_{-\infty}^{\infty} dx\, e^{-i x P_3 z} 
    \int_{-1}^{1} \frac{d\xi}{|\xi|}\, 
    C_{\Delta q}\Big(\frac{x}{\xi},\frac{\mu}{\xi P_3}\Big)\,
    \Delta f_q(\xi,\mu)\ +\ \cdots\, ,
\end{align}
\end{subequations}
where ``$\cdots$'' indicates the neglected power corrections.
The direct matching of the lattice QCD-calculable matrix elements to light-cone PDFs was formally proved to all orders in perturbation theory when $z$ is sufficiently small, with power corrections proportional to
    ${\cal O}(z^2 \Lambda_{\mbox{\tiny QCD}}^2)$
\cite{Ma:2014jla, Ma:2017pxb, Izubuchi:2018srq}.
The relations in Eqs.~(\ref{e.M}) allow us to formally connect the lattice observables to the light-cone PDFs in the same form as the relations connecting experimental observables to the PDFs.
For example, the experimental cross sections for inclusive unpolarized ($\sigma_{\mbox{\tiny DIS}}$) and polarized ($\Delta \sigma_{\mbox{\tiny DIS}}$) DIS and for Drell-Yan (DY) lepton-pair production ($\sigma_{\mbox{\tiny DY}}$) considered in this analysis can be written generically in terms of the spin-averaged $f_i$ and spin-dependent $\Delta f_i$ PDFs as
\begin{subequations}
\label{e.exp}
\begin{align}
\sigma_{\mbox{\tiny DIS}}(x_B,Q^2) 
&= \sum_i \big[ H_{\mbox{\tiny DIS}}^i \otimes f_i
          \big](x_B,Q^2),        \\
\Delta\sigma_{\mbox{\tiny DIS}}(x_B,Q^2) 
&= \sum_i \big[ H_{\mbox{\tiny $\Delta$DIS}}^i \otimes \Delta f_i
          \big](x_B,Q^2),        \\
\sigma_{\mbox{\tiny DY}}(x_F,Q^2) 
&= \sum_{i,j} \big[ H_{\mbox{\tiny DY}}^{ij} \otimes f_i \otimes f_j 
             \big](x_F,Q^2), 
\end{align}
\end{subequations}
where the functions ``$H$'' are the process-dependent, short-distance partonic cross sections, computed up to next-to-leading order (NLO) in perturbative QCD, and here the flavor labels $i$ run over quarks, antiquarks and gluons.
The symbol ``$\otimes$'' represents the convolution integral
$[a \otimes b](x) = \int_x^1 (d\xi/\xi)\, a(x/\xi)\, b(\xi)$.
In Eqs.~(\ref{e.exp}) the variables $x_B$ and $x_F$ are the Bjorken and Feynman scaling variables, respectively, and $Q^2$ is the mass squared of the exchanged or produced virtual boson.
The similarity of the relations between the lattice matrix elements and the PDFs in Eqs.~(\ref{e.M}) and between the experimental cross sections and PDFs in Eqs.~(\ref{e.exp}) allows both types of ``observables'' to be analysed simultaneously within the same global QCD analysis.

While the flavor sums in Eqs.~(\ref{e.M}) and (\ref{e.exp}) run over all partons in the nucleon, the sensitivity of each of the observables to individual parton flavors is different.
In the case of inclusive DIS, the data are mostly sensitive to the sums of quark and antiquark distributions, 
and the use of proton and deuterium data allows one to discriminate between the $u$- and $d$-quark flavors~\cite{Benvenuti:1989rh, Benvenuti:1989fm, Whitlow:1991uw, Arneodo:1996qe, Arneodo:1996kd}.
On the other hand, fixed target DY data from $pp$ and $pd$ collisions offer direct sensitivity to products of quark and antiquark distributions, and careful choice of kinematics can isolate the antiquark distributions in the proton, and in particular its flavor dependence~\cite{Hawker:1998ty, Towell:2001nh}.

For the polarized inclusive DIS case, the constraints from the current data on the spin-dependent PDFs are somewhat weaker, due to the lack kinematic coverage across $x_B$ and $Q^2$, as well as the larger uncertainties on spin-dependent observables.  
In addition, the valence quark number and momentum sum rules that exist for the spin-averaged PDFs are absent for spin-dependent distributions, and only the moments of the isovector and SU(3) octet combinations of spin PDFs can be related with empirical axial vector charges.
The use of flavor SU(3) symmetry in global analyses has been challenged, however, in recent studies of spin-dependent parton distributions~\cite{JAM17}.

To assess the constraints that the lattice data can place on the individual quark and antiquark distributions, we write the light-cone PDFs (which are defined for positive and negative parton momentum fraction $x$) in terms of quark and antiquark components defined with support in the range $x \in [0,1]$.
Using the cross symmetry properties of the spin-averaged and spin-dependent PDFs, we can formally write
\begin{subequations}
\begin{align}
f_q(x)
&= q(x)\, \Theta(x)\Theta(1-x) 
 - \bar{q}(-x)\, \Theta(-x)\Theta(1+x),         \label{e.fq}     \\
\Delta f_q(x)
&= \Delta q(x)\, \Theta(x)\Theta(1-x) 
 + \Delta \bar{q}(-x)\,  \Theta(-x)\Theta(1+x), \label{e.fDq}
\end{align}
\end{subequations}
with each of the quark and antiquark PDFs $q$, $\bar{q}$, $\Delta q$ and $\Delta \bar{q}$ defined for $0 \leq x \leq 1$.
At leading order, quark PDFs and quasi-PDFs are identical, which implies that PDFs are given by the Fourier transform of the matrix elements of Eq.~(\ref{e.ME}).
Considering the real and imaginary parts of the matrix elements separately, one then finds at this order for the spin-averaged distributions the relations
\begin{subequations}
\begin{align}
{\rm Re}\, {\cal M}_q(z,\mu)
&= -\int_0^1 dy\, \cos(y P_3 z)\, \big[ q(y)-\bar q(y) \big] + {\cal O}(\alpha_s^2), \\ 
{\rm Im}\, {\cal M}_q(z,\mu)
&= \int_0^1 dy\, \sin(y P_3 z)\, \big[ q(y)+\bar q(y) \big] + {\cal O}(\alpha_s^2),
\end{align}
\end{subequations}
and for the spin-dependent case,
\begin{subequations}
\begin{align}
{\rm Re}\, {\cal M}_{\Delta q}(z,\mu)
&= -\int_0^1 dy\,\cos(y P_3 z)\, \big[ \Delta q(y) + \Delta \bar q(y) \big] + {\cal O}(\alpha_s^2), \\ 
{\rm Im}\, {\cal M}_{\Delta q}(z,\mu)
&= \int_0^1 dy\,\sin(y P_3 z)\, \big[ \Delta q(y) - \Delta \bar q(y) \big] + {\cal O}(\alpha_s^2),
\end{align}
\end{subequations}
where the higher order corrections are indicated explicitly.
The real part of the matrix element for the unpolarized case is therefore sensitive only to the valence quark PDF, while the imaginary part contains additional information on the antiquark content.
The opposite is true for the spin-dependent distributions, with the imaginary part determined by the polarized valence distributions, and the real part sensitive to the polarization of antiquarks.
The sensitivity of the spin-averaged and spin-dependent sea quark distributions is thus maximal for the imaginary and real parts of their corresponding matrix elements, respectively, and minimal for the conjugate counterparts.

\section{Bayesian framework}
\label{s.bayesian}

Our Bayesian inference for the PDFs is based on the MC methods developed by the JAM Collaboration~\cite{JAM15, JAM17, JAM19}, which involve parametrizing the PDFs at an input scale $\mu_0$ (chosen to be the charm quark mass, $\mu_0=m_c$) and solving the DGLAP evolution equations to evaluate the various observable considered in the analysis at different scales. 
The confidence region for the PDF parameters is characterized by the Bayesian posterior distribution 
\begin{align}
\rho(\bm{a}|{\rm data}) \sim {\cal L}(\bm{a},{\rm data})\, \pi(\bm{a}),
\label{e.bayes}
\end{align}
where $\bm{a}$ represent a vector of the PDF shape parameters, and the likelihood function is chosen to be Gaussian in the $\chi^2$,
\begin{align}
{\cal L}(\bm{a},{\rm data})
    = \exp\Big[-\frac{1}{2}\chi^2(\bm{a},{\rm data})\Big].
\end{align}
The prior function $\pi(\bm{a})$ is flat across the parameter space, and set to zero in regions that give unphysical PDFs.

The MC strategy to sample the posterior distribution utilizes the data resampling approach.
It consist in carrying out a large number of $\chi^2$ minimizations until statistical convergence is achieved.
Each minimization consists of distorting the original data with Gaussian noise, within the quoted uncertainties, and selecting random initial guesses across the region of parameter space allowed by the prior as the starting points for the minimization.
The procedure allows one to scan thoroughly the parameter space, resulting in ensembles of parameters are that distributed according to the posterior distribution.

For the parametrization of the PDFs at the input scale $\mu_0$ we use the traditional form in the literature, given by a generic template shape
\begin{equation}
\label{e.template}
{\rm T}(x,\mu_0^2;\bm{a})
= \frac{a_0}{{\cal N}(\bm{a})}\, x^{a_1}(1-x)^{a_2},
\end{equation}
where the normalization ${\cal N}(\bm{a}) = \int_0^1 dz\, z^{a_1+1}(1-z)^{a_2}$ is defined to be second moment of the distribution in order to decouple the factor $a_0$ from the exponents $a_1$ and $a_2$, and to guarantee a finite normalization for the PDFs.
Using the template shape \eref{template}, for the unpolarized PDFs in the proton we parametrize the valence $u_v \equiv u - \bar u$ and $d_v \equiv d - \bar d$ distributions, the sea-quark $\bar{d}$, $\bar{u}$ and $s = \bar{s}$ distribituions, and the gluon PDF $g$.
The normalization parameters for $u_v$ and $d_v$ are fixed to satisfy the valence number sum rules, while the normalization of the gluon is set by the momentum sum rule.
For the helicity PDFs, we also parameterize the polarized valence $\Delta u_v$ and $\Delta d_v$ distributions, along with $\Delta \bar{d}$, $\Delta \bar{u}$, $\Delta s = \Delta\bar{s}$ and $\Delta g$ using the same template shape \eref{template}.
The normalization parameters for the polarized valence $\Delta u_v$ and $\Delta d_v$ PDFs are determined from the triplet and octet axial charges assuming SU(3) symmetry.

The $\chi^2$ function to be used in the likelihood function is modeled as  
\begin{align}
\chi^2(\bm{a},{\rm data})   
&= \sum_{e,i}
    \left( \frac{d_{e,i} - \sum_k r_{e,k} \beta_{e,k,i} - t_{e,i}(\bm{a})/N_e}
                {\alpha_i}
    \right)^2
 + \sum_k r_{e,k}^2 + \bigg( \frac{1-N_e}{\delta N_e} \bigg)^2,
\end{align}
where $e,i$ label different data sets and data points from different experiments as well as the lattice data, $d_{e,i}$ are the measured observables, and $t_{e,i}$ are the corresponding theoretical values that depend on the shape parameters $\bm{a}$.
The quantity $\alpha_i$ denotes the uncorrelated uncertainties added in quadrature, while $\beta_{k,i}$ are the $k$-th source of point-by-point correlated systematic uncertainties.
The parameters $r_{e,k}$ are nuisance parameters that allow distortion of the theory values additively within the systematic uncertainties $\beta_{k,i}$, and we add a Gaussian penalty as a regulator for these parameters to avoid over-fitting.
While these parameters can be fitted along with the shape parameters $\bm{a}$, they can also be optimized by solving the relation
    $\partial \chi^2(\bm{a},{\rm data})/\partial r_{e,k} = 0$. 
Finally, the theoretical values are modified multiplicatively by the parameter $N_e$, which is included as part of the fitting parameters with a Gaussian penalty of width $\delta N_e$ as a regulator.

The resulting parametrization for the spin-averaged PDFs involves 18 shape parameters and 9 normalization parameters for the experimental data, making a total of 27 free parameters to be inferred using the posterior distribution. 
Similarly, for the spin-dependent PDFs we have 18 shape parameters and 13 data normalizations, for a total of 31 free parameters.

Due to the high dimensionality of the parameter space for the MC sampling of the posterior distribution, we use the multi-step strategy developed in Ref.~\cite{JAM19}.
This allows one to bootstrap efficiently into the region of parameters that have higher likelihood values by performing a sequence of steps to build the MC ensemble of parameters. 
Specifically, we start from flat priors (within the hyperbox) as guess parameters to be optimized by a subset of the data sets.
At each new step, the resulting parameters from the previous step are used as starting parameters for the next iteration, in which the data space is enlarged to include an additional data set.
The sequence is terminated once all the data sets have been included.
This strategy allows one to optimize the parameter ranges efficiently, which would otherwise make the scan of the high dimensional parameter space very inefficient due to the computational cost of evaluating the $\chi^2$ function.

Once the ensemble of parameters $\{\bm{a}\}$ is obtained from the MC sampling, we can estimate expectation values and variances for any PDF or observable ${\cal O}$ computed from the PDFs, using
\begin{subequations}
\begin{align}
{\rm E}[{\cal O}]&=\frac{1}{N}\sum_j^N {\cal O}(\bm{a}_j),      \\   
{\rm V}[{\cal O}]&=\frac{1}{N}\sum_j^N \left({\cal O}(\bm{a}_j)
                   -{\rm E}[{\cal O}]\right)^2.
\end{align}
\end{subequations}
With these analysis tools and theoretical framework outlined in Sec.~\ref{s.methodology}, we can proceed to the numerical analysis of the experimental and lattice data in the next section.

\section{Numerical analysis}
\label{s.analysis}

We begin the discussion of our numerical results with a brief overview of the experimental and lattice data sets used in the our simultaneous global fit.
Following this, we present the results of the analysis for the isovector combinations of the unpolarized and polarized PDFs, and assess the impact of the lattice data on the global PDF extractions.

\subsection{Data sets}

To constrain the spin-averaged PDFs, we use unpolarized inclusive DIS data on proton and deuteron targets from the
    BCDMS~\cite{Benvenuti:1989rh, Benvenuti:1989fm}, 
    SLAC~\cite{Whitlow:1991uw} and
    NMC~\cite{Arneodo:1996qe, Arneodo:1996kd}
experiments, and $e^\pm$-proton cross sections from the
    HERA collider~\cite{Abramowicz:2015mha}.
To provide better flavor separation between the $\bar u$ and $\bar d$ sea quark distributions in the proton, we also utilize the fixed-target $pp$ and $pd$ Drell-Yan data from the E866 experiment at Fermilab~\cite{Hawker:1998ty, Towell:2001nh}.
A total of 2,930 experimental data points are used to constrain the unpolarized PDFs.
For the spin-dependent PDFs, we use inclusive polarized DIS data from the EMC~\cite{EMC89}, 
    SMC~\cite{SMC98, SMC99} and
    COMPASS~\cite{COMPASS10, COMPASS07, COMPASS16}
experiments at CERN, from measurements at
    SLAC~\cite{SLAC-E130, SLAC-E142, SLAC-E143, SLAC-E154, SLAC-E155p, SLAC-E155d, SLAC-E155_A2pd, SLAC-E155x},
and from the HERMES Collaboration~\cite{HERMES97, HERMES07, HERMES12} at DESY.
These total approximately 650 data points for the polarized PDF analysis.
More detailed descriptions of the experimental data can be found in Refs.~\cite{JAM15, JAM17, JAM19, JMO13}.

The lattice data used in this analysis were taken from the simulations by the ETM Collaboration in Ref.~\cite{Alexandrou:2018pbm}, referred to as ``ETMC19.''
The calculation was performed using gauge field configurations at the physical point of two dynamical and degenerate light quarks, on a lattice volume of $48^3 \times 96$, and with a lattice spacing of $a = 0.0938(2)(3)$~fm.
With these parameters the spatial lattice extent was $L \approx 4.5$~fm, with $m_\pi L = 2.98$.
To study the ground state dominance and excited-state contamination, the matrix elements ${\cal M}(z)$ were obtained at four distinct values of the source-sink separations, $t_s/a = 8, 9, 10$ and 12, corresponding to the range $t_s = 0.75 - 1.13$~fm. 
Several analysis methods, including the single-state, two-state and summation methods, were applied to ensure ground-state dominance, as discussed in Ref.~\cite{Alexandrou:2019lfo}. 
The excited state contributions were found to be suppressed for the results with $t_s = 12 a$, and the data were compatible, within uncertainties, with those from the other analysis methods considered.
Three values of the longitudinal momentum $P_3$ were explored, $P_3 = 0.83$, 1.11 and 1.38~GeV; however, to control statistical uncertainties in this analysis we use only the results at the largest momentum, $P_3 = 1.38$~GeV, which were obtained with 72,990 measurements. 
Finally, nucleon mass corrections were taken into account using the results of Ref.~\cite{Chen:2016utp}.

The lattice data are typically given within a symmetric range of $z/a$, between $-15$ and $+15$. 
Strictly speaking, the data points at negative $z$ are fully correlated with those at positive $z$ due the symmetry properties of the matrix elements.
While one can argue that only half of the data points represents the actual information collected by the lattice simulations, in our analysis we include the full spectrum from negative to positive $z$ values in order to impose the symmetry property of the matrix elements as an additional constraint for the Bayesian inference.

For a meaningful estimate of the impact of the lattice data on the global PDFs fits, we perform two separate likelihood analyses, with and without lattice data, for both unpolarized and polarized PDFs.
In principle, an analysis with lattice data alone could also be performed.
However, in practice we find that the PDFs extracted using from lattice data only are not well constrained, and depend rather strongly on the choice of the prior function that constrains the parameter space. 
In particular, PDFs in the small-$x$ region are severely affected by the absence of precision lattice data at large $|z|$ in coordinate space.
While in principle larger values of $|z|$ may constrain PDFs at small $x$, the relations in Eqs.~(\ref{e.M}) themselves break down at large $|z|$ due the presence of large power corrections and the sensitivity to the exponentially $z$-dependent renormalization constant.
As one can see from (\ref{e.M}), however, it is the matrix elements with larger values of $P_3\, z$ that are actually needed in order to access small-$x$ PDFs, which corresponds to lattice configurations with smaller lattice spacing.

This situation is analogous to extracting PDFs from inclusive DIS, where the sensitivity of the data to small-$x$ PDFs is controlled by the ratio $Q^2/s$.
Here, the collision energy squared $s$ and the virtuality of the exchanged photon $Q^2$ play roles analogous to the $P_3$ and $z$ parameters in the lattice matrix element calculation.
Namely, for a fixed value of $s$, one can access small-$x$ PDFs by lowering the values of $Q^2$. 
To access even smaller $x$ values, one needs to increase the collision energy $s$, while keeping $Q^2$ not too low so as to remain in the region where the leading power factorization is valid. 
For the lattice data case, to access smaller values of $x$ one needs to increase $P_3$ but keep $|z|$ not too large. 
The inclusion of the experimental data, on the other hand, helps to constrain precisely those regions of small $x$ that are not yet accessible by lattice data, making the combined global analysis of lattice and experiment data a necessity to maximally utilize the lattice results.

\newpage
\subsection{Unpolarized PDFs}

\begin{table}[b]
\caption{Spin-averaged experimental and lattice QCD data, together with the number of data points and the $\chi^2$ per datum values for the experiment-only and experiment + lattice scenarios.\\}
\small
\begin{tabular}{ l r c c }  \hline
~Observable & {\small \# data}~~~~ & \multicolumn{2}{c}{$\chi^2$/{\small datum}}\\
            & {\small points}~~~~  &  ~~{\small exp}~~ & ~~{\small exp+lat}~~    \\ \hline
~BCDMS $F_2^p$~\cite{Benvenuti:1989rh}  & 348~~~~~                     & 1.1       & 1.1   \\
~BCDMS $F_2^d$~\cite{Benvenuti:1989fm}  & 254~~~~~                    & 1.1       & 1.2   \\
~SLAC $F_2^p$~\cite{Whitlow:1991uw}     & 218~~~~~                     & 1.4       & 1.4   \\
~SLAC $F_2^d$~\cite{Whitlow:1991uw}     & 228~~~~~                     & 1.0       & 1.1   \\
~NMC $F_2^p$~\cite{Arneodo:1996qe}      & 273~~~~~                     & 1.9       & 1.9   \\
~NMC $F_2^d/F_2^p$~\cite{Arneodo:1996kd}& 174~~~~~                     & 1.1       & 1.2   \\
~HERA $\sigma_{\mbox{{\tiny NC}}}^{e^+ p}$ {\small (1)}~\cite{Abramowicz:2015mha}
                                        & 402~~~~~                     & 1.6       & 1.6   \\
~HERA $\sigma_{\mbox{{\tiny NC}}}^{e^+ p}$ {\small (2)}~\cite{Abramowicz:2015mha}
                                        & 75~~~~~                     & 1.2       & 1.2  \\
~HERA $\sigma_{\mbox{{\tiny NC}}}^{e^+ p}$ {\small (3)}~\cite{Abramowicz:2015mha}
                                        & 259~~~~~                     & 1.0       & 1.0 \\
~HERA $\sigma_{\mbox{{\tiny NC}}}^{e^+ p}$ {\small (4)}~\cite{Abramowicz:2015mha}
                                        & 209~~~~~                      & 1.1       & 1.1 \\
~HERA $\sigma_{\mbox{{\tiny NC}}}^{e^- p}$~\cite{Abramowicz:2015mha}
                                        & 159~~~~~                     & 1.7       & 1.7   \\
~HERA $\sigma_{\mbox{{\tiny CC}}}^{e^+ p}$~\cite{Abramowicz:2015mha}
                                        & 39~~~~~                        & 1.4       & 1.2    \\
~HERA $\sigma_{\mbox{{\tiny CC}}}^{e^- p}$~\cite{Abramowicz:2015mha}
                                        & 42~~~~~                        & 1.4       & 1.4    \\
~E866 $\sigma_{\mbox{{\tiny DY}}}^{pp}$~\cite{Hawker:1998ty}
                                        & 121~~~~~                          & 1.3       & 1.3   \\
~E866 $\sigma_{\mbox{{\tiny DY}}}^{pd}$~\cite{Hawker:1998ty}
                                        & 129~~~~~                              & 1.7       & 1.8   \\ 
~ETMC19 ${\rm Re}\, {\cal M}_{u-d}$~\cite{Alexandrou:2018pbm}~~~~~
                                        & 31~~~~~                               &           & 4.7    \\
~ETMC19 ${\rm Im}\, {\cal M}_{u-d}$~\cite{Alexandrou:2018pbm}~~~~~
                                        & 30~~~~~                               &           &22.7~    \\ \hline
~{\bf Total~(exp)}
                                        & 2,930~~~~~                            & 1.3       & ---  \\ 
~~~{\bf ~~~~~~(exp+lat)}~~~
                                        & 2,991~~~~~                            & ---       & 1.6  \\  \hline
\end{tabular}
\label{t.upol}
\end{table}

\begin{figure*}[t]
\centering
\includegraphics[width=1\textwidth]{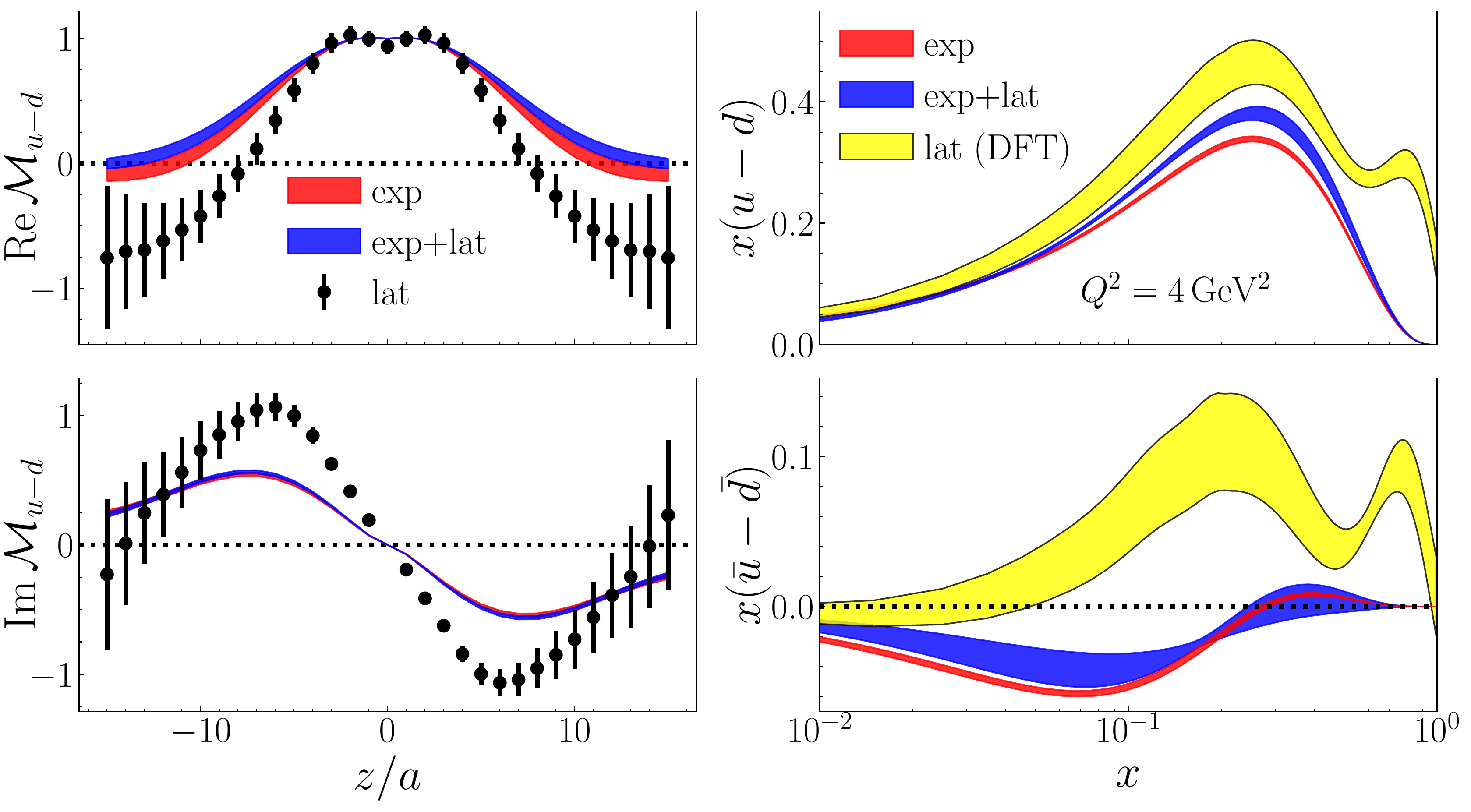}
\caption{Real and imaginary parts of the isovector unpolarized matrix element, ${\cal M}_{u-d}$, in coordinate space ({\it left panels}) and the resulting $x(u-d)$ and $x(\bar u-\bar d)$ PDFs in momentum space ({\it right panels}) at a scale $Q^2=4$~GeV$^2$. The lattice data (black circles) are compared with the fits on the experimental data alone (red bands) and experimental + lattice data (blue bands). The result of the DFT of the lattice data as in Ref.~\cite{Alexandrou:2019lfo} is shown for comparison (yellow bands).}
\label{f.unpol}
\end{figure*}

A summary of the fit results for the spin-averaged PDF analysis is presented in \tref{upol}, where the $\chi^2$ values for each data set are given, along with the total, for the experiment-only and experiment + lattice scenarios.
The fit to the 2,930 inclusive DIS and Drell-Yan experimental data points gives a $\chi^2$ per datum of $\approx 1.3$.
This is a little higher than in recent start-of-the-art global PDF analyses \cite{JAM19, CJ15, MMHT, ABMP, NNPDF, CT18}, mostly because of the very simple functional form chosen in Eq.~(\ref{e.template}).
However, for the purposes of our current analysis, which is to assess the impact of the lattice data on global QCD fits rather than to provide a new set of precision PDFs, the fit is acceptable.

The lattice QCD data provide an additional 61 data points, which are very difficult to fit simultaneously with the experimental points, as reflected in the rather large $\chi^2$ values of 4.7 and 22.7 per datum for the real and imaginary parts of ${\cal M}_{u-d}$, respectively.
The effect of including the lattice data is to increase the total $\chi^2$ by over $\approx 800$ units, leading to an overall $\chi^2$ per datum of 1.6 for the combined experiment + lattice fit.
The current lattice data therefore do not have enough constraining power, relative to the precision of the experimental data, to significantly affect, or improve, the overall global fit of the unpolarized PDFs.

The fit results are illustrated graphically in \fref{unpol}, where we compare the real and imaginary parts of the matrix element ${\cal M}_{u-d}$ from the lattice simulations \cite{Alexandrou:2019lfo}, as a function of $z/a$, with the global fits to the experimental data only and with the combined experiment + lattice analysis results.
For the real part of the matrix element, which is sensitive to the isovector valence quark combination $u_v - d_v$, we find relatively good agreement in the small-$|z|$ region, which reflects the consistency between the lattice and experimental data for the valence number sum rules.
It is also related to the fact that the matching in Eq.~(\ref{e.M}) is valid when $z$ is sufficiently small, whereas the agreement deteriorates at large $|z|$, where the fitted theory prefers a faster suppression with increasing $|z|$.

The imaginary part of the isovector matrix element, which is sensitive to the antiquark asymmetry $\bar u-\bar d$, has an even stronger tension with the experimental data, as seen in the comparison of the lattice values in Fig.~\ref{f.unpol} with the experimental and combined fits.
The discrepancy is dramatically illustrated in the comparison of the isovector combinations of fitted PDFs in \fref{unpol} with the distributions extracted from the direct DFT approach.
For the net $u-d$ distribution, one observes a moderate compatibility of the PDFs with and without lattice data in the valence quark region, but a striking difference in the very high-$x$ region compared with the direct DTF results.
For the isovector antiquark distribution $\bar{u} - \bar{d}$, inclusion of the lattice data does not change the negative sign of the asymmetry at $x \lesssim 0.2$, which is driven largely by the E866 Drell-Yan data~\cite{Hawker:1998ty, Towell:2001nh}, in contrast to the lattice-only extraction via DFT which displays an excess of $\bar u$ over $\bar d$ at intermediate and large $x$ values.

In addition to questions about the validity of factorization at large values of $z$ in Eqs.~(\ref{e.M}), another potential source of the observed tensions is the fact that the renormalization for large $|z|$ is typically of ${\cal O}(100)$, while for small $|z|$ it is ${\cal O}(1)$, which is an effect stemming from the power-law divergence of the Wilson line. 
In addition, the bare matrix element does not decay fast enough, and its effect is enlarged after folding in the renormalization factor, 
preventing a faster decay of the matrix elements in the large-$|z|$ region.
This discrepancy observed in the $x$-dependent PDFs has been discussed in Refs.~\cite{Alexandrou:2018pbm, Alexandrou:2019lfo}.

\subsection{Polarized PDFs}

\begin{table}[b]
\caption{Spin-dependent experimental and lattice data, together with the number of data points and the $\chi^2$ per datum values for the experiment-only and experiment + lattice scenarios.\\}
\small
\begin{tabular}{ l r c c }  \hline
~Observable & {\small \# data}~~  & \multicolumn{2}{c}{$\chi^2$/{\small datum}}\\
            & ~~{\small points}~~  & ~~{\small exp}~~ & ~~{\small exp+lat}~~    \\ \hline
~EMC $A_1^p$~\cite{EMC89}                       & 10~~~~~       & 0.3    & 0.3 \\
~SMC $A_1^p$~\cite{SMC98}                       & 11~~~~~       & 0.6    & 0.7 \\
~SMC $A_1^d$~\cite{SMC98}                       & 11~~~~~       & 2.4    & 2.3 \\
~SMC $A_1^p$~\cite{SMC99}                       & 7~~~~~        & 1.3    & 1.3 \\
~SMC $A_1^d$~\cite{SMC99}                       & 7~~~~~        & 0.7    & 0.7 \\
~COMPASS $A_1^p$~\cite{COMPASS10}               & 11~~~~~       & 1.0    & 0.9 \\
~COMPASS $A_1^d$~\cite{COMPASS07}               & 11~~~~~       & 0.5    & 0.5 \\
~COMPASS $A_1^p$~\cite{COMPASS16}               & 35~~~~~       & 1.0    & 1.0 \\
~SLAC E80/E130 $A_\parallel^p$~\cite{SLAC-E130} & 10~~~~~       & 0.8    & 0.8 \\
~SLAC E143 $A_\parallel^p$~\cite{SLAC-E143}     & 39~~~~~       & 0.9    & 0.8 \\
~SLAC E143 $A_\parallel^d$~\cite{SLAC-E143}     & 39~~~~~       & 1.0    & 1.0 \\
~SLAC E143 $A_\perp^p$~\cite{SLAC-E143}         & 33~~~~~       & 1.0    & 1.0 \\
~SLAC E143 $A_\perp^d$~\cite{SLAC-E143}         & 33~~~~~       & 1.2    & 1.2 \\
~SLAC E155 $A_\parallel^p$~\cite{SLAC-E155p}    & 59~~~~~       & 1.5    & 1.4 \\
~SLAC E155 $A_\parallel^p$~\cite{SLAC-E155d}    & 59~~~~~       & 1.1    & 1.1 \\
~SLAC E155 $A_\perp^p$~\cite{SLAC-E155_A2pd}    & 46~~~~~       & 0.8    & 0.8 \\
~SLAC E155 $A_\perp^d$~\cite{SLAC-E155_A2pd}    & 46~~~~~       & 1.5    & 1.5 \\
~SLAC E155x $\tilde{A}_\perp^p$~\cite{SLAC-E155x} & 69~~~~~     & 1.3    & 1.3 \\
~SLAC E155x $\tilde{A}_\perp^d$~\cite{SLAC-E155x} & 69~~~~~     & 0.9    & 0.9 \\
~HERMES $A_1^n$~\cite{HERMES97}                 & 5~~~~~        & 0.3    & 0.3 \\
~HERMES $A_1^p$~\cite{HERMES07}                 & 16~~~~~       & 0.6    & 0.6 \\
~HERMES $A_1^p$~\cite{HERMES07}                 & 16~~~~~       & 1.3    & 1.3 \\
~HERMES $A_2^p$~\cite{HERMES12}                 & 9~~~~~        & 1.1    & 1.1 \\
~ETMC19 ${\rm Re}\, {\cal M}_{\Delta u-\Delta d}$~\cite{Alexandrou:2018pbm}~~~ & 31~~~~~       &        & 0.5 \\
~ETMC19 ${\rm Im}\, {\cal M}_{\Delta u-\Delta d}$~\cite{Alexandrou:2018pbm}~~~ & 30~~~~~       &        & 0.3 \\
\hline
~{\bf Total~(exp)}
                                    & 651~~~~~ & 1.1       & ---  \\ 
~~~~~~~~~{\bf ~(exp+lat)}~~~
                                    & 712~~~~~ & ---       & 1.0  \\  \hline
\end{tabular}
\label{t.pol}
\end{table}

\begin{figure*}[t]
\centering
\includegraphics[width=1\textwidth]{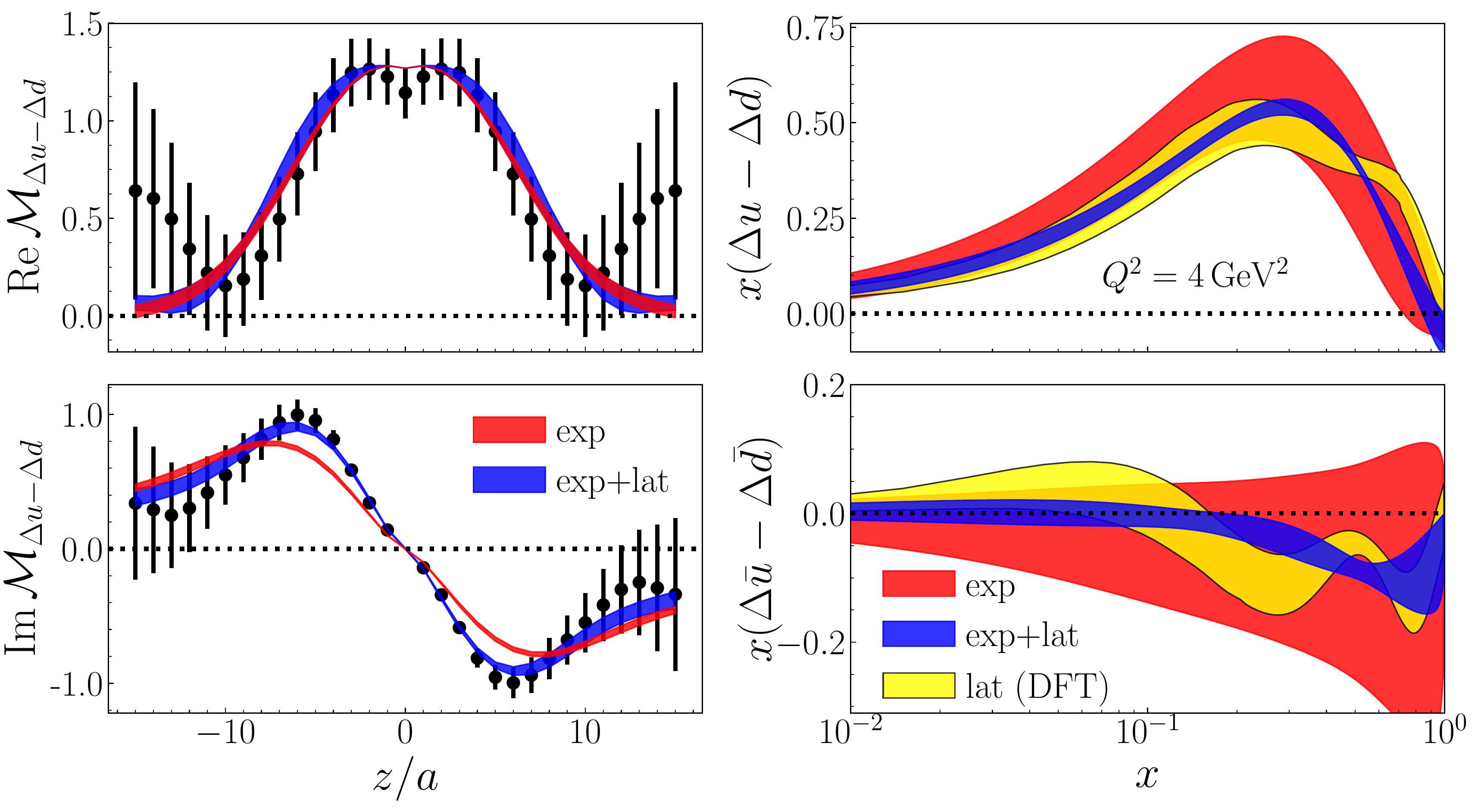}
\caption{As in Fig.~\ref{f.unpol}, but for the spin-dependent isovector matrix element, ${\cal M}_{\Delta u -\Delta d}$, and PDFs $x(\Delta u-\Delta d)$ and $x(\Delta\bar u-\Delta\bar d)$.}
\label{f.pol}
\end{figure*}

The results for the spin-dependent PDF analysis, in which 651 polarized inclusive DIS data points are fitted together with an additional 61 lattice data points, are shown in \tref{pol}.
The total $\chi^2$ per datum for the experiment-only fit is 1.1, and actually decreases slightly to 1.0 when the lattice data are added, in contrast to what was found in the unpolarized analysis above.
The smaller values of the $\chi^2$ per datum, $\sim 0.5$, for the lattice data reflects both the larger uncertainties on the spin-dependent lattice matrix elements and the closer general correspondence between the shapes of the lattice and empirical coordinate space distributions.

The level of agreement between the lattice and experimental data can be seen in \fref{pol} for the spin-dependent coordinate space matrix element ${\cal M}_{\Delta u-\Delta d}$. 
For the real part of the matrix element, which is sensitive to the isovector combination $\Delta u + \Delta\bar u - \Delta d - \Delta\bar d$, we find excellent agreement in the small-$|z|$ region, indicating the consistency of the lattice triplet axial charge with experiment.
The agreement deteriorates at larger values of $|z|$, where the tails of the matrix element increasingly deviate away from zero, but are still consistent within the large uncertainties with the fit to experimental data.

The predictions for the imaginary part, which is sensitive to the spin-dependent isovector valence combination $\Delta u_v - \Delta d_v$, have the largest discrepancies at intermediate values of $z/a$ if one uses only the experimental data.
On the other hand, the combined experiment + lattice data analysis allows an optimal solution to be found that simultaneously describes both types of data sets.
The fact that the lattice data are able to change the posterior distribution indicates that already the existing lattice data can provide significant constraints on spin-dependent PDFs beyond that from the existing experimental measurements.

The impact of the lattice data on the spin-dependent PDF uncertainties is illustrated in \fref{pol}, where the uncertainties on the PDFs fitted to experimental data alone are significantly reduced when supplemented with the lattice matrix element data.
Also shown in \fref{pol} are the PDFs inferred via the DTF method, which are found to be in relatively good agreement with the PDFs extracted using the continuum approach within current uncertainties.
In contrast to the spin-averaged case in \fref{unpol}, however, we find that for the spin-dependent PDFs the lattice data do not suggest any large $\Delta \bar u - \Delta \bar d$ asymmetry, as suggested by the data on the longitudinal single-spin asymmetry for $W^\pm$ production from RHIC~\cite{Adam:2018bam}.
More accurate lattice data and better matching coefficients could help improve the determination of this asymmetry.

\subsection{Moments}

\begin{figure*}[t]
\centering
\includegraphics[width=1\textwidth]{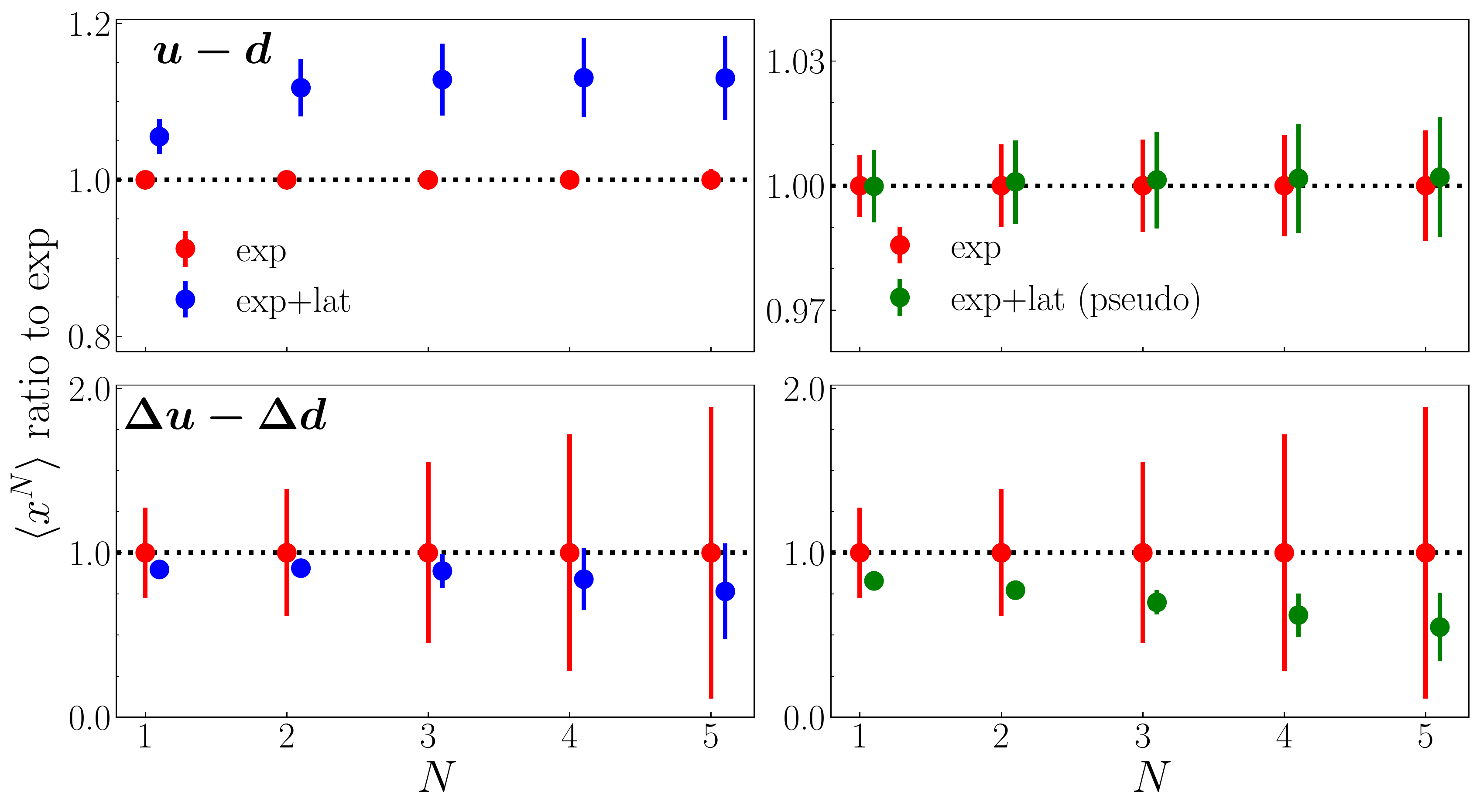}
\caption{Ratios of $N$-th moments $\langle x^N \rangle$ of the isovector unpolarized $u-d$ ({\it top row}) and polarized $\Delta u-\Delta d$ ({\it bottom row}) PDFs from the experiment-only (red circles), experiment plus lattice (blue circles), and experiment with pseudo-lattice data (green circles) fits, relative to the experiment-only results.}
\label{f.moments}
\end{figure*}

To further test the quality of the agreement between the lattice and experimental data, we consider moments of PDFs, which are less sensitive to point-to-point statistical fluctuations of the PDFs in momentum space.
The $N$-th moments of the spin-averaged ($q$) and spin-dependent ($\Delta q$) PDFs in Eqs.~(\ref{e.fq}) and (\ref{e.fDq}) are defined by
\begin{subequations}
\begin{align}
    \left<x^N\right>_q
    &= \int_{-1}^1 dx\, x^N f_q(x),     \\
    \left<x^N\right>_{\Delta q}
    &= \int_{-1}^1 dx\, x^N \Delta f_q(x),
\end{align}
\end{subequations}
respectively.
In \fref{moments} the first five moments are compared relative to the moments extracted from experimental data alone. 
For the spin-averaged case, a clear pattern of disagreement is observed when the experimental are fitted together with the lattice matrix elements.
Although the lattice-constrained results differ from the experiment-only fits by $\sim$10\%--15\%, because of the small uncertainties on the data this represents a discrepancy of up to $\approx$~4$\sigma$.
For the spin-dependent case, on the other hand, the moments are generally in very good agreement when extracted with or without the lattice data, and indicate a significant reduction of uncertainties when the lattice data are included in the analysis.

Assuming that the issues that generate the disagreement between lattice and experimental data for the spin-averaged case are resolved, a relevant question to ask is whether the current precision of the lattice calculations can have an impact on the spin-averaged PDFs. 
To answer this question, we carry out a pseudo-data analysis using as ``lattice'' data the central predictions from the PDFs extracted from experimental data alone, but keeping the actual uncertainties quoted by the lattice calculations.
As can be seen from \fref{moments}, the impact of the current precision of the lattice data on spin-averaged PDFs is marginal.
In contrast, the reduction of the uncertainties on the moments for the spin-dependent case is quite significant.

\begin{figure*}[t]
\centering
\includegraphics[width=1\textwidth,keepaspectratio]{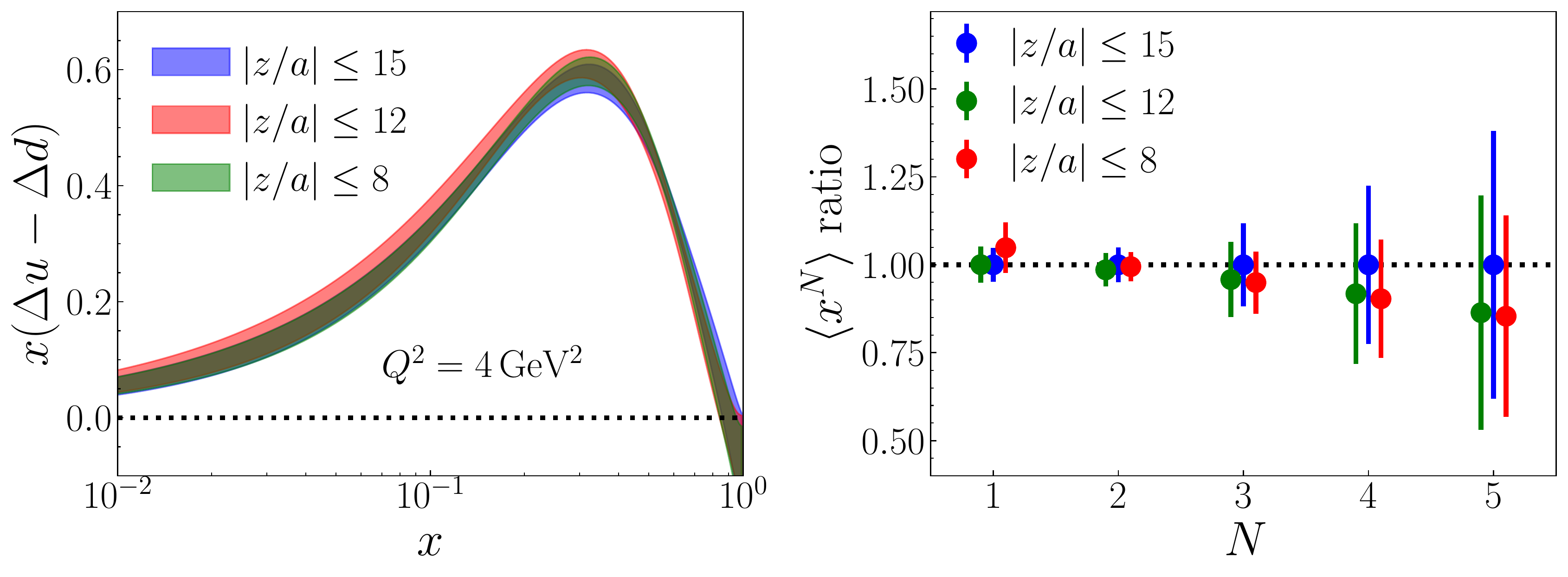}
\caption{Effects of the cuts on $z/a$ on the spin-dependent isovector PDF $x(\Delta u-\Delta d)$ ({\it left}) and on its $N$-th moments $\langle x^N \rangle$ relative to the full result with $|z/a| \leq 15$ ({\it right}), for the cut values $z_{\rm max}/a = 8$ (green band and circles), 12 (red), and 15 (blue).}
\label{f.ppdf-cuts}
\end{figure*}

Focusing on the spin-dependent PDFs, we also explore the importance of the large-$|z|$ lattice matrix elements, which generally have larger statistical errors than the low-$|z|$ data.
The larger uncertainties stem from the renormalization factor growing exponentially with $|z|$ due to power-law divergences, and greater sensitivity to power corrections and lattice systematic uncertainties.
To address this question we perform several additional fits using subsets of the lattice data with cuts on the maximum values of $|z|$ used in the analysis.
The results in \fref{ppdf-cuts} generally indicate stability of the spin-dependent PDFs with respect to varying the cut between $z_{\rm max}/a = 15$, which includes all of the computed lattice data, and $z_{\rm max}/a = 8$, which excludes most of the tails in which the errors on the matrix elements grow large.
The extracted spin-dependent PDFs from each of the regions considered are within the estimated errors, and the ratios of the moments for $z_{\rm max}/a = 8$ and 12 to the full results with $z_{\rm max}/a = 15$ are unity within the quoted uncertainties.

\section{Conclusions}
\label{s.conclusions}

In this paper we have performed the first combined MC-based global QCD analysis of spin-averaged and spin-dependent PDFs, using both experimental data and matrix elements computed from first principles in lattice QCD.
Utilizing the recent calculations of the isovector matrix elements from Ref.~\cite{Alexandrou:2018pbm}, we combined these in our likelihood analysis on the same footing as the experimental data using the continuous Fourier transform approach relating matrix elements with the PDFs.

For the spin-averaged PDFs, we found significant tension between the lattice and experimental data, with the precision of the latter dominating the inference on the PDFs. 
The precision of the unpolarized lattice matrix elements will need to improve significantly in future for the lattice data to be competitive with experiment in constraining PDFs.
In contrast, our analysis of the spin-dependent PDFs shows promising agreement between the lattice and experimental data, with the current precision of the polarized matrix elements providing significant constraints on PDFs compared with the precision of existing experimental data. 
In particular, we find that the existing lattice results do not indicate a large $\Delta \bar{u}-\Delta \bar{d}$ asymmetry, as suggested in recent data on $W$ production in single-spin asymmetries observed at RHIC~\cite{Adam:2018bam}.

Since the direct matching between lattice calculable matrix elements and PDFs in Eq.~(\ref{e.M}) has been shown to be valid to all orders in QCD perturbation theory when  $z$ is sufficiently small, in the future more precise data from lattice configurations with smaller lattice spacing will be needed.
This will allow for a larger range of hadron momentum $P_3$ and values of $z$, which will provide better constraints on the $x$ dependence of PDFs, especially in the spin-dependent sector.
Our analysis demonstrates the feasibility of including lattice data in standard QCD global analysis, and paves the way for future precision studies of PDFs that can treat information from experimental data and lattice QCD on equal footing.

\section*{Acknowledgements}

This work was supported by the DOE contract No.~DE-AC05-06OR23177, under which Jefferson Science Associates, LLC operates Jefferson Lab, and DOE contract No.~DE-SC0008791.
J.B. was supported by the DOE CSGF program, Award No.~DE-SC0019323.
M.C. acknowledges support by the DOE, Office of Science, Office of Nuclear Physics, within the framework of the TMD Topical Collaboration. F.S.\ was funded by DFG project number 392578569.
The work of N.S. was supported by the DOE, Office of Science, Office of Nuclear Physics in the Early Career Program.


\end{document}